\documentclass[iop,apj]{emulateapj}
\usepackage{amsmath,amssymb,amstext}

\usepackage[breaklinks,colorlinks,citecolor=blue,linkcolor=magenta]{hyperref} 

\usepackage[all]{hypcap} 
\usepackage{xfrac}

\usepackage{amssymb}
\usepackage{amsmath}
\usepackage{graphicx}

\newcommand{\bnabla}{{\mbox{\boldmath$\nabla$}}}

\usepackage{aas_macros}
\usepackage{natbib}
\shorttitle{Powering of H$\alpha$ filaments}
\shortauthors{Ruszkowski, Yang, and Reynolds}

\begin{document}

\title{Powering of H$\alpha$ filaments by cosmic rays}
\author{Mateusz Ruszkowski$^{1,2}$, H.-Y. Karen Yang$^{2,3,5}$, and Christopher S. Reynolds$^{2,3,4}$} 
% * <mateuszr@umich.edu> 2017-12-12T18:48:13.894Z:
%
% ^.
\affil{$^{1}$Department of Astronomy, University of Michigan, 1085 S University Ave, 311 West Hall, Ann Arbor, MI 48109}
\affil{$^{2}$Department of Astronomy, University of Maryland, College Park, MD 20742}
\affil{$^{3}$Joint Space-Science Institute (JSI), College Park, MD 20742}
\affil{$^{4}$Institute of Astronomy, University of Cambridge, Madingley Road, Cambridge, CB3 0HA, UK}
\altaffiltext{5}{{\it Einstein} Fellow}
\email{mateuszr@umich.edu (MR), hsyang@astro.umd.edu (KY), chris@astro.umd.edu (CR)}

\begin{abstract}
Cluster cool cores possess networks of line-emitting filaments. These filaments are thought to originate via uplift of cold gas from cluster centers by buoyant active galactic nuclei (AGN) bubbles, or via local thermal instability in the hot intracluster medium (ICM). Therefore, the filaments are either the signatures of AGN feedback or feeding of supermassive black holes.  Despite being characterized by very short cooling times, the filaments are significant H$\alpha$ emitters, which suggests that some process continuously powers these structures. Many cool cores host diffuse radio mini halos and AGN injecting radio plasma, suggesting that cosmic rays (CRs) and magnetic fields are present in the ICM. We argue that the excitation of Alfv{\'e}n waves by CR streaming, and the replenishment of CR energy via accretion onto the filaments of high plasma-$\beta$ ICM characterized by low CR pressure support, can provide the adequate amount of heating to power and sustain the emission from these filaments. This mechanism does not require the CRs to penetrate the filaments even if the filaments are magnetically isolated from the ambient ICM and it may operate irrespectively of whether the filaments are dredged up from the center or form in situ in the ICM. This picture is qualitatively consistent with non-thermal line ratios seen in the cold filaments. Future X-ray observations of the iron line complex with {\it XARM}, {\it Lynx}, or {\it Athena} could help to test this model by providing constraints on the amount of CRs in the hot plasma that is cooling and accreting onto the filaments.
\end{abstract}
 
\keywords{galaxies: clusters: intracluster medium -- cosmic rays}
\maketitle

\section{Introduction}
H$\alpha$ filaments are ubiquitous in cool core galaxy clusters and giant elliptical galaxies (e.g., \citet{McDonald2010,Werner2014}). These filaments are known to contain multiphase gas emitting CO lines (e.g., \citet{Salome2006}), far-infrared lines \citep{Werner2013}, near-infrared emission lines (e.g., \citet{Donahue2000}), optical lines (e.g., \citet{Canning2011}), and soft X-rays \citep{Fabian2003}. They are also likely to be significantly magnetized \citep{Fabian2008}. \\
\indent
Two leading hypotheses for the origin of these structures suggest that they formed as a result of either dredging up of the cold gas from the very centers of the potential wells of ellipticals and clusters by the AGN (e.g., \citet{Fabian2003}), or that they formed in situ in the atmospheres of these objects as a result of precipitation and are in the process of accretion (e.g., \citet{Kunz2012}, \citet{McCourt2012, Gaspari2012, Li2016}). Thus, the filaments most likely represent the signatures of either AGN feedback or feeding of the central supermassive black holes in clusters and ellipticals \citep{li2017}.\\
\indent
A typical cooling timescale of the H$\alpha$ filaments is much shorter than the buoyancy timescale or the dynamical time in cool cores. This suggests that the filaments need to be continuously powered or they will become invisible in the optical band. This is also consistent with the fact that most filaments do not form stars \citep{Canning2010, Canning2014}. Several heating processes have been suggested to explain the filament powering. Photoionization by the central AGN has been proposed as the solution (e.g., \citet{Heckman1989}) but, at least in the Perseus cluster, it can be ruled out as the primary heating process due to the fact that the H$\alpha$ luminosity does not decrease with the distance from the cluster center \citep{JohnstoneFabian1988}. Photoionization by massive stars has been proposed but the observed line ratios are in conflict with those seen in HII regions \citep{KentSargent1979} and many of the filaments do not have enough star formation \citep{canning2016}. Shock heating has been proposed \citep{SabraShields2000,farage2010} but can be ruled out as the general mechanism because this model overpredicts the observed OIII line emission \citep{VoitDonahue1997} and leads to tension between the expected and observed dependence of line ratios on the gas velocity dispersion \citep{Canning2011}. X-ray heating of the filaments by the hot ambient ICM has been ruled out by the high H$_{2}/$H$\alpha$ ratios \citep{Donahue2000}. Thermal conduction of heat from the ambient medium into the filaments has been considered \citep{Voit2008} as a promissing mechanism to power the filaments but it is possible that conduction may be severely limited by plasma processes \citep{Komarov2014, Komarov2016, RobergClark2016}. Other meachnisms have been proposed and include reconnection of magnetic fields in the filaments uplifted by buoyant AGN bubbles \citep{Churazov2013}, excitation in turbulent mixing layers \citep{CrawfordFabian1992,BegelmanFabian1990}, and heating due to collisions with the energetic particles surrounding filaments \citep{Ferland2008,Ferland2009,Donahue2011}.\\
\indent
Radio mini halos -- diffuse radio sources comparable in size to the scales of cluster cores -- have been observed in many galaxy clusters. Recent results show that the radio mini halos are present in many cool core clusters and absent from non-cool core ones \citep{giacintucci2017}. Due to the proximity of the Virgo cluster, observations of this object reveal that, while the radio emission is coincident with the buoyant bubbles, it is also clearly detected in the bulk of the ICM \citep{deGasperin2012}. In Perseus radio emission is coincident with the AGN bubbles but it is seen to extend beyond the bubbles at lower radio frequencies \citep{Fabian2002}. These observations are consistent with a possibility that CRs diffuse out of the AGN bubbles into the ambient ICM \citep{ruszkowski2008}. In general, the elongated filamentary cold-gas structures are immersed within the regions that are radio emitting in both the Virgo and Perseus clusters. Interestingly, the spatial distribution of radio emission in Perseus is not uniform and tends to correlate with the position of the Northern filament \citep{gendron2017}. While the radio emission comes from relativistic CR electrons, associated with these electrons should be energetic CR protons. \citet{DunnFabian2004} suggest that the magnetic pressure and the pressure of the radio-emitting CR electrons inside the bubbles is small compared to the ambient ICM pressure. This suggests that the bubbles are filled with either non-radiating CR particles or ultra-hot thermal gas. Consequently, the diffusion of these particles out of the AGN bubbles leads to the interaction of CRs with the ambient ICM. CR streaming along magnetic fields excite waves and Alfv{\'e}n wave heating was proposed as a viable mechanism to offset radiative cooling in global one-dimensional simulations of cluster atmospheres \citep{Loewenstein1991,GuoOh2008, Fujita2011,Fujita2012, Fujita2013, Pfrommer2013, JacobPfrommer2017a,JacobPfrommer2017b}. Two-dimensional simulations of local thermal instability with adiabatic CR component \citep{SharmaParrishQuataert2010} demonstrated that  the filaments may be CR-pressure-dominated. Most recently \citet{Ruszkowski2017} included the effect of CR streaming instability heating in three-dimensional magnetohydrodynamical (MHD) simulations of AGN feedback in clusters. The above arguments suggest that CRs are an important component within the cool cores and could carry valuable information about the interplay between the radiative cooling, AGN feedback, and the formation of the filaments. 
\\
\indent
Here we propose that H$\alpha$ filaments can be powered by Alfv{\'e}n wave heating associated with the streaming of CRs along the magnetic fields inside the filaments. In our model CRs and magnetic fields are amplified in the filaments as a result of accretion and cooling of ambient ICM onto the filaments. This mechanism should operate irrespectively of whether the filaments are dragged out of the cluster centers or formed in situ via thermal instability. Also, it does not require particles external to the filaments to penetrate them to supply adequate heating nor does it require high non-thermal pressure support in the bulk of the ICM.\\

\section{Powering of filaments by cosmic rays}

\subsection{Dissipation of CR energy by streaming instability}

\begin{figure}
  \begin{center}
    \leavevmode
        \includegraphics[width=0.45\textwidth]{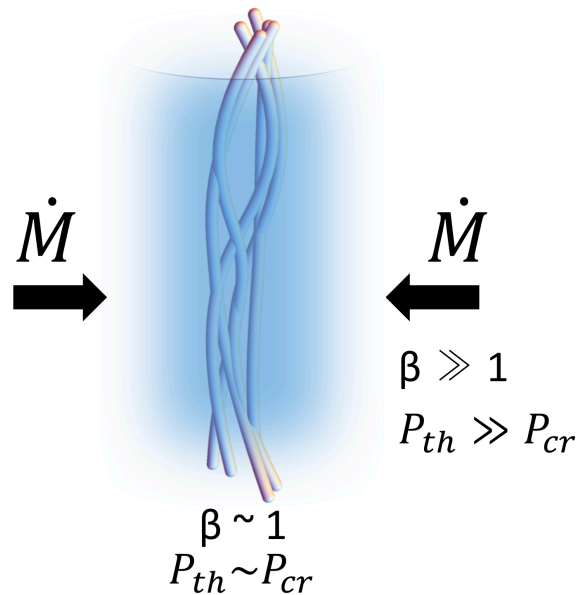}       
       \caption[]{H$\alpha$ filament consisting of a number of strands. Radiative cooling leads to accretion of high-plasma-$\beta$ ICM characterized by low CR pressure support. Non-thermal pressure support inside the filament is significant and creates conditions favorable for CR heating via streaming instability. }
     \label{fig:dens}
  \end{center}
\end{figure}

In the self-confinement model \citep{skilling1971,zweibel2013}, CRs stream down their pressure gradient along the magnetic fields and scatter on self-induced magnetic field perturbations due to the so-called streaming instability. In the process, CRs experience an effective drag force and heat the thermal gas at the rate of $-{\bf v}_{A}\cdot\bnabla P_{\rm cr}$, where ${\bf v}_{A}={\bf B}/\sqrt{4\pi\rho}$ is the Alfv{\'e}n velocity\footnote{Note that the Alfv{\'e}n speed in partially ionized medium will be larger by $1/\sqrt{x}$, where $x$ is the ionization fraction. However, since H$\alpha$ emission requires a significant fraction of the gas to be ionized, we ignore this term.} and $P_{\rm cr}$ is the CR pressure. 
The efficiency of the heating due to the streaming instability $\hat{\epsilon}_{\rm si}$ can be defined as
\begin{align}
\hat{\epsilon}_{\rm si} \equiv \frac{\frac{r}{2}(-{\bf v}_{A}\cdot\bnabla P_{\rm cr})}{c_{s0} n k_{b}T}, \label{eq1}  
\end{align}
where the nominator is the amount of energy generated due to the streaming instability per surface area of the filament per time, or the filament energy flux, $c_{s0}$ is the sound speed in the ambient ICM, $T$ is the ambient temperature, $n$ is the total particle number density, and $r$ the filament radius. Our objective is to quantify $\hat{\epsilon}_{\rm si}$ and compare it directly to the constraints on flux from the H$\alpha$-emitting filaments in the Perseus and Virgo clusters.\footnote{In Section 2.2 we discuss the coupling of CR to the weakly ionized filament gas and argue that CR are well coupled and can deposit energy at the rate given by Eg.\ref{eq1} even in the presence of ion-neutral friction damping of the self-excited Alfv{\'e}n waves.}\\
\indent
The magnetic field in the filaments is likely to consist of a component parallel to the length of the filaments and a turbulent component. Thus, in general we can decompose the field as ${\bf B} = {\bf B}_{||} + {\bf B}_{\perp}$, where  ${\bf B}_{||}$ and ${\bf B}_{\perp}$ are the components parallel and perpendicular to the main axis of the filament, respectively. In the process of filament formation the magnetic field is likely significantly amplified beyond its value in the ambient ICM. As the cross section of the filament decreases, $B_{||}\propto r^{-2}\propto\rho^{2/3}$. Thus, if the filament initially consists of predominantly parallel component, e.g., due to the vertical stretching of the fields in the wake of a buoyantly rising bubble, then the total strength of the field $B\propto\rho^{2/3}$. In the process of bubble rise, or during infall of filaments formed in situ in the ICM, shearing motions and Kelvin-Helmholtz instability operating on the interface between the filaments and the ambient ICM will likely generate $B_{\perp}$ very close to the filaments. The timescale for this instability to develop is
\begin{align}
t_{\rm KH}\sim\omega_{\rm KH}^{-1}\sim 0.6\: r_{30}\Delta v_{200}^{-1}  \left(    \frac{     \sqrt{   \rho_{\rm fil}  /  \rho_{\rm icm} } } {20} \right){\rm Myr},
\end{align}
where $\Delta v_{200}=\Delta v/200$km/s is the velocity difference between the filament and the ambient ICM, $r_{30}=r/30$pc is the filament radius and $\rho_{\rm fil}$ and $\rho_{\rm icm}$ are the filament and ICM densities, respectively. Consequently, even for quite conservative parameter choices, there is ample time for this instability to tangle the magnetic fields during the bubble rise time 
$t_{\rm buoy}\sim L_{3}  \Delta v_{200}^{-1}  \sim 15$ Myr, where $L_{3}=L/3$ kpc and $L$ is the filament length and $\Delta v_{200}=\Delta v/200$ km s$^{-1}$, or during the filament infall. 
Magnetic field tension may somewhat slow down the development of this instability. Alternatively, the filament could initially contain random fields. In this case, $B_{\perp}\propto r^{-1}$ and $B_{||}\propto r^{-2}$, so the filament collapse would preferentially amplify $B_{||}\propto\rho^{2/3}$. Kevin-Helmholtz and other plasma instabilities (firehose, mirror, or the streaming instability itself), could then again generate $B_{\perp}$. \\
\indent
We now assume that some fraction $f$ of the magnetic energy density will be channeled to the $B_{||}$ component and the remaining energy density to $B_{\perp}$. Thus, we partition the energy such that 
\begin{align}
B=B_{0}\left(\frac{\rho}{\rho_{0}}\right)^{2/3}; \frac{B^{2}_{\perp}}{8\pi}=(1-f)\frac{B^{2}}{8\pi};\; \frac{B^{2}_{||}}{8\pi}=f\frac{B^{2}}{8\pi}. 
\end{align}
Using this notation we can write the volume heating rate as
\begin{equation}
\begin{split}
-{\bf v}_{A}\cdot\bnabla P_{\rm cr} & \sim 2 v_{A,x}\frac{\partial P_{\rm cr}}{\partial x} + v_{A||}\frac{\partial P_{\rm cr}}{\partial z}\\ 
 & \sim 2(1-f)^{1/2}\frac{B}{\sqrt{4\pi\rho}}\frac{P_{\rm cr}}{r},\label{eq3} 
\end{split}
\end{equation}
where we neglected the heating associated with the CR gradient along the filament. The filament is oriented along the $z$-axis and the $x$-axis is perpendicular to it. 
The heating rate is limited by the maximum magnetic and CR pressures in the filament. These pressures are bounded by the pressure of the ambient ICM. Assuming that a fraction of the filament pressure support $f_{B}$ can be attributed to the magnetic pressure, this argument leads to the magnetic field strength 
of $B\sim B_{0}\beta^{1/2}f_{B}^{1/2}$, where the plasma $\beta$ parameter $\beta=P_{\rm icm}/P_{B, {\rm icm}}$, where $P_{\rm icm}$ and $P_{B, {\rm icm}}$ are the thermal and magnetic pressures in the ambient ICM. \\
\indent
Note that the limit on the magnetic pressure translates into the critical density $\sim\rho_{0}\beta^{3/4}f_{B}^{3/4}$, where $\rho_{0}$ is the ambient ICM density in the vicinity of the filament, beyond which the magnetic pressure would exceed the thermal pressure of the ambient ICM. Similarly, we can introduce a fraction $f_{\rm cr}$ of CR pressure to the total pressure in the filament to find a critical filament density that corresponds to the balance of CR pressure in the filament and the ambient ICM pressure. For example, since the observed density of the H$\alpha$-emitting phase in the filaments is larger than the above estimate of the critical density (e.g., for $\beta\sim 100$, $f_{B}\sim 0.5$, and $\rho_{0}$, the critical density is $\sim 2$cm$^{-3}$, which is likely below the density of the H$\alpha$-emitting phase in the filaments; see below), this suggests that the magnetic fields (and CRs) either partially dissipated their energy or partially leaked out in the process of the filament formation. Nevertheless, as mentioned below, the filament pressure support is likely to be significantly affected by these non-thermal components.\\
\indent
Given that both the CR and magnetic pressures scale as $\rho^{\gamma_{\rm cr}}$ in the initial stages of the filament formation (where $\gamma_{\rm cr}=4/3$ is the adiabatic index of CRs), the ratios of these pressures is constant $P_{\rm cr}/P_{B}=P_{\rm cr, icm}/P_{B, {\rm icm}}=X_{\rm cr}\beta$, where $X_{\rm cr}$ is the ratio of the ambient CR pressure $P_{\rm cr, icm}$ to the ambient thermal pressure. Thus, for representative values of $X_{\rm cr}\sim 0.3$  for M87 \citep{Pfrommer2013} and $\beta\sim 100$, the CR contribution to the pressure support in cool filaments may be very important even if the CR pressure contribution to the total ICM pressure support is much smaller.
More generally, \citet{JacobPfrommer2017a,JacobPfrommer2017b} argue that typical levels of CR pressure support in cool cores may be around $\sim0.1$ of the ICM pressure. Observations of M87 suggest that the filaments have not cooled down isobarically and that, indeed, a significant pressure component in the $10^{4}$K phase emitting H$\alpha$+[NII] is non-thermal in origin \citep{Werner2013}. Specifically, \citet{Werner2013} show that assuming {\it thermal} pressure balance between the filaments and the ambient ICM implies that the density of the filament H$\alpha$+[NII] phase could be up to $\sim4$ times larger than actually detected. Since the relative contributions of the CR and magnetic pressures to the overall pressure balance are uncertain, we simply parametrize them via $f_{\rm cr}$ and $f_{B}$. Thus, the limiting value of the Alfv{\'e}n speed is
\begin{align}
v_{A}=\frac{{\rm min}\left[B_{0}\left(\frac{\rho}{\rho_{0}}\right)^{2/3}, B\right]}{\sqrt{4\pi\rho}}\sim
\left(\frac{2}{\gamma}\frac{\rho_{0}}{\rho}\right)^{1/2}f_{B}^{1/2}c_{s0} \label{eq4} 
\end{align}
where $\gamma=5/3$ is the adiabatic index of the thermal gas, $\rho_{0}$ is the density of the ambient ICM in the vicinity of the filament and $\rho$ is the density of the H$\alpha$-emitting phase. The last approximate equality in Eq. \ref{eq4} is a consequence of the fact that is easy to amplify the magnetic field to the point where it becomes limited by the ambient ICM pressure. For typical values relevant to M87/Virgo cluster $\rho_{0}\sim 0.1$ cm$^{-3}$, $\rho\sim 30$ cm$^{-3}$, and $T\sim 1.5$ keV  \citep{Werner2013,zhuravleva2014}, we get $v_{A}\sim 40 f_{B}^{1/2}$km s$^{-1}$.\\
\indent
By combining Eqs. \ref{eq1}, \ref{eq3} and \ref{eq4} we can now write the efficiency of the filament emission as
\begin{align}
\hat{\epsilon}_{\rm si}\sim\left(\frac{2}{\gamma}\frac{\rho_{0}}{\rho}\right)^{1/2}(1-f)^{1/2}f_{\rm cr}f_{B}^{1/2}.   \label{eq5} 
\end{align}
The above expression assumes that a filament is volume filling. It has been suggested \citep{fabian2011,Werner2013} that filaments may consist of sub-strands of cold gas. While this suggestion is based in part on the assumption that the filament cooling is isobaric, structures smaller than few tens of pc do remain unresolved by Hubble, which is consistent with the hypothesis that the filaments may consist of a number of sub-filaments corresponding to certain volume filling factor $f_{V}$ and covering factor $f_{A}$. Assuming that a filament consists of $N$ strands each characterized by radius $\lambda$, the total flux through the side surface enclosing the volume occupied by the original volume filling filament is 
\begin{align}
\frac{ N\pi\lambda^{2}L  }{2\pi rL}(-{\bf v}_{A}\cdot\bnabla P_{\rm cr}) \propto
\frac{1}{2r}N\lambda^{2}v_{A\perp}\frac{P_{\rm cr}}{\lambda} \sim 
\frac{f_{A}v_{A\perp}P_{\rm cr}}{2}, 
\label{eq6} 
\end{align}
where the last approximate equality comes from the definition of the surface covering factor $f_{A}\equiv (2\lambda L N)/(2 r L)$, which is the ratio of the sum of the projected side surface area of the strands comprising the filament to the size of the projected side of the original volume filling filament. Using Eqs \ref{eq5} and \ref{eq6} we can now write the efficiency of a filament consisting of a number of strand as
\begin{align}
\epsilon_{\rm si}\sim f_{A}\hat{\epsilon}_{\rm si}\sim\left(\frac{2}{\gamma}\frac{\rho_{0}}{\rho}\right)^{1/2}(1-f)^{1/2}f_{\rm cr}f_{B}^{1/2}f_{A}.   \label{eq7} 
\end{align}
Note that if significant substructure in the filaments is also seen along the major filament axes, then the contribution from CR streaming parallel to the filaments should also be included in Eq. \ref{eq3}. Adding  this component would further increase the expected heating efficiency.\\
\indent
Using M87 data we can now estimate the observed filament efficiency $\epsilon_{\rm M87}$. Assuming that the bolometric flux is 20 times larger than the H$\alpha$ flux, the observed bolometric flux is $2.2\times 10^{-3}$erg  s$^{-1}$cm$^{-2}$ \citep{Werner2013}. From the definition of $\epsilon$, this flux needs to be compared to $\sim \epsilon_{\rm M87} 2 n_{e}k_{B}Tc_{s0}$, where $n_{e}$ and $T$ are the electron number density and gas temperature in the ambient ICM in the vicinity of the filament, respectively. \citet{Werner2013} suggest $\epsilon_{\rm M87}\sim 0.22$, while \citet{Churazov2013} considers $\epsilon_{\rm M87}\sim 0.1$, but the exact value depends on the assumed ambient gas density and temperature which are uncertain. Using representative electron number density $n_{e}\sim 0.1$ cm$^{-3}$ and the gas temperature $T\sim 1.5$ keV \citep{zhuravleva2014}, we obtain $\epsilon_{\rm M87}\sim 0.073$. As an example, in order to reconcile efficiency $\epsilon_{\rm si}\sim 0.1$ with our model, we require only a modest covering factor $f_{A}\sim 7$ for plausible parameter choices $f=1/3$, $f_{B}=1/4$, and $f_{\rm cr}=1/2$. An example set of model parameters that simultaneously satisfies observational constraints is shown in Table \ref{tbl:table1}.

\capstartfalse
\begin{deluxetable*}{ccccccccc}
\tablecaption{Main model parameters$^a$ \label{tbl:table1}}
%\tabletypesize{\scriptsize}
\tablecolumns{8}
\tablewidth{400pt}
\tablehead{
	\multicolumn{2}{c}{Ambient ICM} & \colhead{} & \multicolumn{5}{c}{Filaments} \\
	\cline{1-2} \cline{4-9} \\
	\colhead{$\beta$} & \colhead{X$_{\rm cr}$} & \colhead{} & \colhead{$\epsilon_{\rm si}$} & \colhead{$f_{s}f_{A}$} & \colhead{$f_{\rm cr}$} & \colhead{$f_{B}$} & \colhead{$f$}
}
\startdata
100 & 0.1 		&& 0.1 & 0.66 & 0.5 & 0.25 & 0.33
\enddata
\tablenotetext{a}{From left to right the columns show plasma $\beta$ in the ambient ICM, ratio of CR pressure to the total ICM pressure in the cluster cool core $X_{\rm cr}$, heating efficiency $\epsilon_{\rm si}$, product of the filament covering fraction $f_{A}$ and the ratio $f_{s}$ of the speed of gas accretion onto the filaments to the ambient ICM sound speed, ratio of the CR-to-total pressure in the filaments ($f_{\rm cr}$), ratio of the magnetic-to-total pressure in the filaments ($f_{B}$), and the fraction of magnetic pressure in the component parallel to the filament compared to the total magnetic pressure ($f$).}
\tablecomments{An example set of model parameters that simultaneously satisfies observational constraints (see text for details).}
\end{deluxetable*}

\subsection{Coupling of CRs to the weakly-ionized filament gas}
Coupling of CRs to gas demands that CRs scatter sufficiently frequently on the self-excited Alfv{\'e}n waves. At gas temperatures $\sim 10^{4}$K, that correspond to the H$\alpha$-emitting phase, the filament gas will become partially ionized. Under such conditions ion-neutral damping of the CR-excited Alfv{\'e}n waves may become strong and the CR may decouple from the gas and may not deposit their energy \citep{kulsrud1971}. 

In order to assess if this decoupling affects transport of CRs in the H$\alpha$-emitting plasma, we can estimate CR drift speed $u_{s}$ by comparing the streaming instability growth rate \citep{Wentzel1968,kulsrud1969,skilling1971,KulsrudBook} to the ion-neutral damping rate \citep{kulsrud1971}. The growth rate is given by
\begin{equation}
\Gamma = \frac{\pi}{4}\Omega_{o}\frac{n_{\rm cr}}{n_{i}}\frac{n-3}{n-1}\left(\frac{u_{s}}{v_{A,i}}-1\right), 
\end{equation}
where $\Omega_{o}=eB/m_{p}c$ is the non-relativistic gyrofrequency, $n_{\rm cr}$ and $n_{i}$ are the number density of CRs and ions, respectively, $n$ is the slope of the CR distribution function in momentum, and $v_{A,i}=v_{A}/\sqrt{x}$, and $x$ is the ionization fraction. The damping rate is  
\begin{equation}
\Gamma_{\rm in}=\frac{\nu_{\rm in}}{2}=5.1\times 10^{-9}T_{4}^{1/2}n_{n}
\end{equation}
where $n_{n}$ is the number density of neutrals and where the ion-neutral collision frequency $\nu_{\rm in}$ was  computed using the expressions from \citet{dePontieu2001}. Equating $\Gamma$ to $\Gamma_{\rm in}$ leads to 
\begin{equation}
\frac{u_{s}}{v_{A}}=\left[1 + 1.5\times 10^{-6}T^{1/2}_{4}(1-x)x\frac{n_{H}^{2}}{B_{\mu}n_{\rm cr}} \right]\frac{1}{\sqrt{x}},\label{stream_in} 
\end{equation}
where $T_{4}=T_{\rm fil}/10^{4}$K, $n_{H}$ is the hydrogen number density, and $B_{\mu}=B/\mu$G with all parameters corresponding to the filament gas. The CR number density inside the filaments can be estimated by assuming that CRs contribute a fraction $f_{\rm cr}$ of the total filament pressure
\begin{equation}
n_{\rm cr}=3\times 10^{-7}\frac{n-4}{n-3}f_{\rm cr,-1}n_{e,-1}T_{\rm icm, 5}E_{\rm min, GeV}^{-1},
\end{equation}
where $f_{\rm cr,-1}=f_{\rm cr}/0.1$, $n_{e,-1}=n_{e}/0.1$cm$^{-3}$ and $T_{\rm icm, 5}=T_{\rm icm}/5$ keV are the ion number density and the temperature of the ambient ICM, and $E_{\rm min, GeV}=E_{\rm min}/$GeV is the minimum energy of the CR distribution. Similarly, magnetic field can be estimated by assuming that the magnetic pressure in the filament is a fraction $f_{B}$ of the ICM pressure, i.e., $B=\sqrt{8\pi f_{B}P_{\rm icm}}$. Using parameter values representative of M87 ($n_{H}=$ 30 cm$^{-3}$, $T_{4}=1$, $n_{e,-1}=1$, and $T_{5}=0.3$), and assuming as above that $f_{\rm cr}=1/2$, $f_{B}=1/4$, and that $n=4.6$ and $E_{\rm min,GeV}=1$, and using a conservative value of $x=0.5$, the term is square brackets in Eq. \ref{stream_in} that corresponds to ion-neutral damping evaluates to $\sim 43$, and $u_{s}\gg v_{A}$. This suggests that CRs begin to decouple from the gas as they enter the low ionization phase of the filaments. 

However, the fact that CR transport is faster in the cold filaments does not immediately imply that the CR heating due to the streaming instability vanishes. As long as the mean free path of CRs remains smaller than the filament thickness, CRs can still  scatter on self-excited waves.
Note that the faster CR drift speed does not imply boosted heating rates (e.g., \citet{RYZ17}). As long as the self-trapping of CRs takes place, CRs 
deposit energy in the ICM according to Eq. \ref{eq1}. Following \citet{WienerZweibelOh2013} we estimate CR mean free path $\lambda_{\rm cr}$ in the regime where ion-neutral damping dominates and compare $\lambda_{\rm cr}$ to the filament width $L_{\rm cr}$
\begin{equation}
\frac{\lambda_{\rm cr}}{L_{\rm cr}}\sim 5\times 10^{-12}(n-1)x^{1/2}(1-x)\frac{n^{3/2}_{H}}{n_{\rm cr}}\left(\frac{p}{p_{c}}\right)^{n-3}, \label{eq13} 
\end{equation}
where $p$ is the CR momentum and $p_{c}$ is the low momentum cutoff of the CR spectrum. Eq. \ref{eq13} is valid for relativistic CRs, and we assume that typical values of $p_{c} \sim$ GeV/c. \citet{WienerZweibelOh2013} use $p/p_{c}=100$. However, a representative value of $p/p_{c}$ is the ratio of the CR momentum corresponding to the average CR energy $\langle E\rangle$ to the cutoff momentum, i.e., $p(\langle E\rangle)/p_{c}\sim 3$, which depends weakly on the minimum CR energy. Thus, for $n_{H}=30$ cm$^{-3}$, $n_{\rm cr}=1.5\times 10^{-7}$cm$^{-3}$, and $x=0.5$ (which corresponds to the values corresponding used in the previous paragraph), we get $\lambda_{\rm cr}/L_{\rm cr}\sim 4\times 10^{-2}$, which implies that CRs are still trapped and can deposit their energy inside the filaments. 

\subsection{Resupply of energy}
The energy radiated away by the filaments needs to be resupplied to them at the same rate. This energy could be replenished either by allowing CRs to escape along the wake of the rising bubbles where the filaments may be preferentially located \citep{ruszkowski2008} or by accretion onto the filaments of new gas from the tenuous phase of the ICM. It can be demonstrated that while the fraction of bubble energy that would need to leak out into the wake region for this balance to occur is much smaller than unity, the timescale over which this energy is radiated away as the CRs travel down the wake is significantly shorter than the bubble rise time or the timescale over which the CRs could propagate along the filaments. Therefore, we instead turn our attention to the latter idea where CR energy is supplied by accretion of ambient ICM onto the filaments. In this case, the above timescale issue is naturally circumvented because CRs are delivered at all locations along the filament simultaneously. However, the issue of whether sufficient power is supplied to the filaments needs to be addressed. Additional advantage of this mode of energy resupply is that the filaments need not be located in the filaments trailing behind buoyant AGN bubbles. \\
\indent
The timescale over which the CR energy is lost to heating of the gas inside the filaments is the ratio of the energy per surface area surrounding the entire bundle of subfilaments to the energy flux corresponding to that surface area
\begin{align}
t_{\rm heat}\sim \frac{f_{\rm cr}P_{\rm icm}}{\gamma_{\rm cr} -1}\frac{\pi r^{2}Lf_{V}}{2\pi rL}\frac{1}{\hat{\epsilon}_{\rm si}c_{s0}P_{\rm icm}f_{A}}.  \label{eq8} 
\end{align}
Therefore, the ratio of the CR power
\begin{align}
P_{\rm heat}\sim \frac{P_{\rm icm}Vf_{V}f_{\rm cr}}{\frac{1}{2}\frac{r}{c_{s0}}\frac{f_{V}}{f_{A}}f_{\rm cr}   }    {\hat{\epsilon}_{\rm si} }\label{eq9} 
\end{align}
to the power supplied by accretion onto the region containing all subfilaments of the ICM gas containing CRs
\begin{align}
P_{\rm supply}\sim 2\pi rLf_{A}P_{\rm cr, icm}v_{r}  \label{eq10} 
\end{align}
is
\begin{align}
\frac{P_{\rm supply}}{P_{\rm heat}}\sim  \frac{f_{A}X_{\rm cr}} {\epsilon_{\rm si}} \frac{v_{r}} {c_{s0}},\label{eq11} 
\end{align}
where $v_{r}$ is the speed of accretion in the direction perpendicular to the filaments. We note that stretching of the filaments along their axes (e.g., due to uplift) results in their narrowing, accretion of ambient ICM, and non-zero $v_{r}$ \citep{Churazov2013}. However, much faster accretion occurs due to fast radiative cooling of the ICM in the vicinity of the filaments. The pressure in the ICM surrounding the filaments is dominated by thermal gas pressure. However, close to the filaments radiative losses are large and the thermal energy is quickly radiated away. When the cooling timescale $t_{\rm cool}\sim k_{b}T/(n_{e}\Lambda)$ becomes shorter than the sound crossing time $t_{\rm sc}\sim r/c_{s}$, the gas is no longer supported by thermal pressure and begins to flow toward the filaments due to pressure gradients at speeds that are no longer negligible compared to the sound speed in the ambient ICM. Thus, from Eq. \ref{eq11}, we see that the rate of energy supply can match the rate at which the energy is dissipated.
Conservative estimate for these timescales for parameters representative of conditions near M87 filaments ($r\sim 30$ pc, $n_{e}\sim 30$ cm$^{-3}$, $T\sim 10^{4}$K, $\Lambda\sim 10^{-24}$erg s$^{-1}$cm$^{3}$) is $t_{\rm cool}\sim 1.5\times 10^{3}$yr and $t_{\rm sc}\sim 2\times 10^{6}$yr. While the exact value of the cooling function near $10^4$K varies significantly with temperature, the accreting ICM needs to transition through the peak in the cooling function near $10^{5}$K at which point the cooling time is even shorter than the above estimate. Thus, cooling essentially tends to be isochoric near the filaments and the accretion onto the filaments begins to accelerate. This observation agrees with the findings of \citet{li2014}, who study internal structure of the cold filaments forming in the ICM due to thermal instability. In particular, they find that the thermal pressure support in the cold phase is significantly lower than in the ambient ICM and that the cold phase accretes from the hot ICM at a significant fraction of the hot phase sound speed albeit with large uncertainties. This leads to an interesting prediction that in a narrow layer surrounding the filaments the gas should be approaching the filaments at a non-negligible fraction of the sound speed in the ambient ICM. Detecting such gas would require very good spectral and spatial resolution in the soft X-rays.\\
\indent
Associated with this CR energy resupply is the mass accretion rate that we estimate to be $\dot{M}\sim 2\pi f_{A}rL\rho_{\rm 0}f_{s}c_{s0}$, where $f_{s}$ is the fraction of the speed of sound of the inflowing material. For $n_{e}\sim 0.1$ cm$^{-3}$, $c_{s0}\sim 625$ km s$^{-1}$, $r\sim 30$ pc, $L\sim5$ kpc, we get $\dot{M}\sim 1.4f_{s}f_{A}$M$_{\odot}$yr$^{-1}$. We note that this accretion rate will be reduced for filaments at larger distances from the cluster center where the ambient ICM density is lower. We also point out that, unlike the CR component of the gas advected close to the filaments, the thermal energy of this gas is low due to very fast cooling and, consequently, mixing in of the thermal component into the filaments does not significantly contribute to the filament powering. We note that our mechanism does not rely on the penetration of the filaments by {\it external} particles in order to provide adequate heating. In our model CRs are simply advected with the magnetized ambient gas that is accreting onto the filaments and these CRs do not need to propagate across the magnetic fields during the process of accretion. 
Furthermore, it is plausible that the magnetic fields in the ambient ICM are less tangled than in the filaments, and if so, in the process of accretion onto the filaments these external fields could shield the filaments and prevent filament evaporation due to thermal conduction even if conduction remains unsuppressed.\\
\indent
We can translate gas accretion rates on individual filaments into the overall cooling rates. In the Virgo cluster the number of filaments appears to be smaller than for the Perseus cluster, so in order to put conservative constraints on the overall cooling rates here we consider the Perseus cluster. Using data for the Northern filament in Perseus we get $\dot{M}\sim 1.9f_{s}f_{A}$M$_{\odot}$yr$^{-1}$, where we assume ambient temperature of $\sim4$ keV, ambient density of 0.04 cm$^{-3}$, and the filament length of 9 kpc and radius of 35 pc. For example, for $f_{s}f_{A}\sim 1/1.5$ and $X_{\rm cr}\sim 0.15$, we can ensure that the supplied and dissipated heat are comparable (c.f. Eq. 17). Using filament H$\alpha$ flux of $7\times 10^{-4}$erg cm$^{-2}$s$^{-1}$ \citep{fabian2011}, we get H$\alpha$ luminosity $\sim 1.3\times 10^{40}$erg s$^{-1}$. We now rescale the mass accretion rate by the ratio of the total H$\alpha$ luminosity $\sim10^{42}$erg s$^{-1}$ to that of the Northern filament, and obtain the total cooling rate $\sim95$ M$_{\odot}$yr$^{-1}$, which is broadly consistent with the observations. In estimating the total H$\alpha$ luminosity based on the results of \citet{Heckman1989}, we excluded the emission from within $\sim$10 kpc away from the cluster center because the elongated filaments, the emission from which is the focus of this paper, are located outside this region. Excluding this emission reduces the H$\alpha$ luminosity by about a factor of two. Note also that any diffuse H$\alpha$ emission remains unexplained by our model as our heating mechanism requires significant CR gradients to be present. Alternatively, following \citet{Ferland2009} we could use H$\alpha$ luminosity of $\sim 7\times 10^{39}$erg s$^{-1}$ for the Horseshoe filament that is approximately two times shorter. Note that the GMOS slit size that is used to extract this luminosity is larger than the filament radius assumed above. However, since HST measurements indicate that the filament is significantly narrower than the slit width, we assume that this emission comes entirely from within this radius. Under these assumption the total cooling rates based on the Horseshoe filament are essentially the same as those for the Northern filament. 

\subsubsection{Advection of CRs onto the filaments}
As mentioned in Section 2.1, some fraction of CR and magnetic energy either partially dissipated or leaked out of the filaments. This is indeed required if the model is to explain why the observed density of the H$\alpha$ gas exceeds the critical density that one would expect if the filaments were supported by the magnetic and CR pressure. Our filament heating model relies on the self-confinement picture of CR transport, where the gas is heated via the streaming instability and transport occurs via streaming. In the extrinsic turbulence model, where the transport occurs via diffusion, CRs do not heat the gas as they are transported down the CR pressure gradients. Thus, by construction, our model does require at least CR streaming is present. In this section we argue that the escape of CRs from the filaments due to CR streaming does not have to be very efficient, but that diffusion (if present) could help to remove some CRs.\\
\indent
Whether CRs can escape the filament depends on the competition between the accretion speed of the ambient gas onto the filaments and the speed at which CRs stream out of the filaments. As argued above, accretion should occur at some fraction of the sound speed in the hot ambient ICM because this ambient gas responds to rapid loss of pressure in the very vicinity of the filament caused by fast cooling of the gas near the peak of the cooling function. Therefore, in the estimates of the accretion and streaming speeds discussed below, we use parameters representative of the conditions in the accreting gas. \\
\indent
In the self-confinement model, CRs stream at the Alfv{\'e}n speed but damping of self-excited waves can increase CR propagation speed. Under conditions relevant to those in the vicinity of the filaments, turbulent damping dominates over non-linear Landau damping. Equating the wave growth rate to the dissipation rate, one can derive the effective CR transport speed (e.g., \citet{wiener2013,Ruszkowski2017})
\begin{equation}
u_{s}=v_{A}\left(1 + 0.08\frac{B_{10}^{1/2}n_{i,-2}^{1/2}}{L_{\rm mhd, 10}^{1/2}   n_{c,-9}} \right),\label{stream} 
\end{equation}
where $n_{i,-2}=n_{i}/10^{-2}{\rm cm}^{-3}$ and $n_{c,-9}=n_{c}/10^{-9}{\rm cm}^{-3}$ are the ion and CR number densities respectively, $B_{10}=B/10\mu\rm{G}$ and 
$L_{\rm mhd, 10}=L_{\rm mhd}/10 {\rm kpc}$ is the scale where turbulence is Alfv{\'e}nic (we assumed that the slope of the CR momentum distribution is 4.6 and the average CR Lorentz factor is 3). Assuming conservatively that $L_{\rm mhd, 10}=0.1$, and using $X_{\rm cr}=0.1$, $n_{i,-2}=10$, minimum energy of CRs of 1 GeV, and the ICM temperature of 1.5 keV, we get $n_{c,-9}\sim 51$. For plasma $\beta\sim 10^{2}$, CR streaming speed is $u_{s}\sim v_{A}$. \\
\indent
In the high-plasma-$\beta$ gas, such as the ICM surrounding the filament, Landau damping can boost the CR propagation speed beyond that expected in the presence of just turbulent damping such that the second term on the right hand side of Eq. \eqref{stream} is multiplied by $\beta^{1/2}\sim 10$ \citep{wiener2018}. This leads to a moderately super-Alfv{\'e}nic CR transport speed $u_{s}\sim 1.2v_{A}$. Given that Alfv{\'e}n speed is a fraction of the sound speed, $v_{A}/c_{s}=[2/(\gamma\beta)]^{1/2}\sim 0.1$, the CR streaming speed can be comparable to the speed of the ICM accreting onto the filament, but it does not necessarily exceed the accretion speed (even when the covering factor $f_{A}>1$; see discussion of mass accretion rate in Section 2.3). \\
\indent
If CR transport occurs via diffusion, then the CR transport speed $\kappa/r\sim 10^{8}$cm s$^{-1}$ (assuming diffusion coefficient $\kappa\sim 10^{28}$cm$^{2}$ s$^{-1}$ and $r\sim 30$ pc) could exceed accretion speed if the latter occurs at a fraction of the ambient ICM sound speed $\sim 5\times 10^{7}f_{s}$cm s$^{-1}$. The average transport of CRs out of the filaments could be reduced if the magnetic field inside the filaments is dominated by the parallel component (large $f$; e.g., due to filament uplift by AGN bubbles or radial infall toward the center). In order to ensure that the filaments are efficiently heated in this case, we require that $(1-f)^{1/2}f_{A}$ remains unchanged (c.f. Eq. \eqref{eq7}). This implies larger covering factors $f_{A}$ and consequently smaller gas accretion speed onto the filaments to ensure that the ratio $P_{\rm supply}/P_{\rm heat}\sim 1$ and that the total mass accretion rate remains consistent with observations (see Section 2.3). Such reduced accretion speed could occur when the gas inflow toward the filaments is slowed by nonthermal pressure support that is required by our model. A reduced accretion speed means that CRs could nevertheless be escaping, and the escape would be faster than in the streaming case discussed above. However, even if nominally the CR escape speed exceeds the accretion speed, the average transport of CRs away from the filaments is not likely to be efficient once CRs reach the regions immediately adjacent to the filaments where the CR pressure gradient will vanish. These CRs will still be subject to preferential advection of the ICM toward the filaments and we speculate that CRs could be mixed into the cold filament gas via instabilities operating on the ICM-filament interface. Nevertheless, as required by our model, some escape of CRs from the filaments could occur in this case. \\

\subsection{Dissipation of CR energy by other mechanisms}
\subsubsection{Turbulent dissipation}
The gas inside the filaments may be turbulent and the dissipation of this turbulence could in principle also contribute to the filament powering. In order to obtain a very rough estimate of the turbulent dissipation rate, we assume that the velocity dispersion $\sigma$ in the filaments is at best comparable to the sound speed in the H$\alpha$-emitting filament gas. Otherwise, there should be evidence for shock heating but that is not observed. In general shock heating should lead to a correlation between [NII]/H$\alpha$ emission line ratios and the velocity dispersion of the gas, but such correlations have not been detected in the Perseus and Centaurus clusters where this issue was studied (\citet{Hatch2006,Canning2011}; albeit the caveat that these models did not incorporate nonthermal pressure that could affect the nature of the shocks). The absence of shocks implies that 
\begin{equation}
\sigma\la \left(1+\frac{\gamma_{\rm cr}f_{\rm cr}+f_{B}}{\gamma f_{g}}\right)^{1/2}c_{s,0},
\end{equation}
where $f_{g}$ is fraction of thermal pressure support in the filament and $c_{s,0}$ is the sound speed in the absence of any nonthermal pressure in the filament for $10^{4}$K. For $f_{g}=f_{\rm cr}=f_{B}=1/3$, we get $\sigma\sim 24$ km s$^{-1}$. The turbulent power $L_{\rm turb}\sim 1.5 M_{\rm fil}\sigma^{3}/l_{\rm turb}$. Typical masses of filaments are in the range from $M_{\rm fil}\sim$10$^{4}$ to 10$^{6}$ M$_{\odot}$\citep{Conselice2001}. Using filament mass of $10^{6}$M$_{\odot}$ and assuming turbulence injection scale comparable to the filament width $l_{\rm turb}=60$ pc (c.f. \citet{canning2016} who use smaller value), we get $L_{\rm turb}\sim2\times 10^{38}$erg $s^{-1}$, which is a few percent of the H$\alpha$ luminosity of the resolved Horseshoe or Northern filaments in Perseus. This is not a strict upper limit on the contribution of turbulent dissipation as the filament may consist of a number of subfilaments. However, inside the filaments the H$\alpha$ phase, while possibly not completely volume-filling, may be more volume-filling than the phase corresponding to the dense molecular gas. Furthermore, the power contributed by turbulent dissipation depends on the uncertain mass in the H$\alpha$ phase in the filament, and that mass is smaller than the total $M_{\rm fil}$ mass of the filament adopted above. Most importantly however, the measurements of internal turbulence in the filaments are very difficult because of limited spatial resolution. Current measurements of the velocity dispersion inside the filaments are very likely to be significantly overestimated due to filament or sub-filament confusion \citep{canning2016} and consequently velocity dispersions could be consistent with values lower than those adopted above. We thus conclude that it is at least plausible that turbulent heating is not the dominant powering mechanism.

\subsubsection{Hadronic, ionization, and Coulomb losses}
In addition to the heating associated with the CR streaming instability other mechanisms may be responsible for the transfer of energy from CRs to the thermal gas. Specifically, CRs will suffer hadronic, ionization, and Coulomb losses. Assuming the energy density in CRs is dominated by protons at $\sim 3$ GeV, the corresponding cooling times can be approximated as \citep{tova2013} $t_{\rm hadron}\sim 1.8\times 10^{8} n_{p}^{-1}$ yr, $t_{\rm ion}\sim 5.2\times 10^{8} n_{n}^{-1}$ yr, and $t_{\rm coulomb}\sim 3.1\times 10^{8} n_{e}^{-1}$ yr, where $n_{n}$ is the number density of neutral medium. Direct observational constraints on the gas density in the H$\alpha$-emitting phase  in M87 can be obtained from [SII]$\lambda$6716/[SII]$\lambda$6731 line ratios \citep{Werner2013} and yield particle density $n_{\rm part}\sim 2n_{e}\sim 30$ cm$^{-3}$. 
If we conservatively assume that $n_{e}\sim 0.5n_{p}$ for these conditions, and also conservatively assume that $n_{n}\sim n_{p}$ and use $n_{p}\sim 30$ cm$^{-3}$, then all of these timescales exceed the heating timescale (c.f. Eq. \ref{eq8}),
\begin{align}
t_{\rm heat}\la \frac{f_{\rm cr}}{2(\gamma_{\rm cr}-1)\epsilon_{\rm si}}\frac{r}{c_{s0}} \sim 10^{6} f_{\rm cr} {\rm yr}, 
\end{align}
for $r\sim 30$ pc, and where the upper limit comes from considering just a single filament. This implies that the CR heating in the H$\alpha$-emitting phase is dominated by that due to the streaming instability. 

\section{Summary and conclusions}
We presented a model for powering of H$\alpha$ filaments by CRs. The main conclusions presented in this paper can be summarized as follows.
\begin{enumerate}
\item We suggest that the CR streaming instability could be a significant contributor to sustained powering of H$\alpha$ filaments in the atmospheres of galaxy clusters and elliptical galaxies. The proposed mechanism offers an alternative to other filament heating mechanisms such as magnetic field reconnection (that may operate in the wakes of rising AGN bubbles \citep{Churazov2013}), excitation of turbulent mixing layers \citep{CrawfordFabian1992}, and heating due to collisions with the energetic particles surrounding filaments (that may require penetration of filaments by energetic particles \citep{Ferland2009}).
\item The proposed mechanism should operate irrespectively of whether the filaments are dredged up by AGN bubbles or form in situ in the ICM via local thermal instability, and it does not rely on the filaments being magnetically connected to the ambient ICM.
\item Heating of the filaments is likely to be significant even if the magnetic and CR pressure support in the bulk of the ICM is very low compared to the thermal ICM pressure.
\end{enumerate}
\acknowledgments{The authors thank the referee for useful comments. M.R. acknowledges NASA ATP 12-ATP12-0017 grant and NSF grant AST 1715140. H.Y.K.Y. acknowledges support from NSF grant AST 1713722, NASA ATP (grant number NNX17AK70G) and the Einstein Postdoctoral Fellowship by NASA (grant number PF4-150129). C.S.R. thanks for the support from the US NSF under grant AST 1333514. C.S.R also thanks NASA for support under grant NNX17AG27G. M.R. thanks Rebecca Canning, Julie Hlavacek-Larrondo, Megan Donahue, Ellen Zweibel, Alberto Bolatto, Jay Gallagher, Richard Mushotzky, and Erin Kara for useful discussions. M.R. thanks Department of Astronomy at the University of Maryland for hospitality during his sabbatical stay. M.R. thanks Suvi Gezari for letting him derive most of the estimates presented in this paper on the windows of her office. M.R. is grateful for the hospitality of the Astronomy Department at the University of Wisconsin--Madison, which was made possible in part by a generous gift from Julie and Jeff Diermeier. This work was performed in part at the Aspen Center for Physics, which is supported by NSF grant PHY-1066293.\\}

\bibliography{filaments}

\end{document}